\documentclass[preprint,12pt]{elsarticle}

\usepackage{amssymb,amsmath,bm}

\def\ii{\int\limits_{\mathbb{R}}}

\journal{Nonlinear Analysis: Real World Applications}

\begin{document}

\begin{frontmatter}



\title{Hamiltonian model for coupled surface and internal waves in the presence of currents}


\author{Rossen Ivanov}

\address{School of Mathematical Sciences, Dublin Institute of Technology,\\ Kevin Street, Dublin 8, Ireland\\
and
\\
Environmental Sustainability and Health Institute (ESHI), Dublin Institute of Technology \\
Grangegorman, Dublin 7, Ireland  }

\begin{abstract}
We examine a two dimensional fluid system consisting of a lower medium bounded underneath by a flatbed and an upper medium with a free surface. The two media are separated by a free common interface. The gravity driven surface and internal water waves (at the common interface between the media) in the presence of a depth-dependent current are studied under certain physical assumptions. Both media are considered incompressible and with prescribed vorticities. Using the Hamiltonian approach the Hamiltonian of the system is constructed in terms of 'wave' variables and the equations of motion are calculated. The resultant equations of motion are then analysed to show that wave-current interaction is influenced only by the current profile in the 'strips' adjacent to the surface and the interface. Small amplitude and long-wave approximations are also presented.
\end{abstract}


\end{frontmatter}


\section{Introduction}\label{}
\noindent{\sc }

It has been known for many centuries that the ocean contains currents that flow along generally consistent paths. The Spanish galleons transporting gold and silver from Mexico to Spain made use of the Gulf Stream to help them return home. Since then, scientists have gained much more information on both where currents flow and why. In the oceans currents very often exist with undercurrents. The first undercurrent was discovered in 1951 by Townsend Cromwell who was investigating fishing techniques in the central Pacific Ocean. Undercurrents have since been found under most major currents.  The equatorial region in the Pacific is characterised by a thin shallow layer of warm and less dense water over a much deeper layer of cold denser water. The two layers are separated by a sharp {\it thermocline} (where the temperature gradient has a maximum, it is very close to the {\it pycnocline}, where the pressure gradient has a maximum)  at a depth, depending on the location, but usually at 100 -- 200 m beneath the surface.  For modelling purposes both layers are assumed homogeneous with a sharp boundary at the thermocline/pycnocline (see \cite{JM}).

The Equatorial Undercurrent (EUC) flows in a region that is roughly within 200 - 300 km (below 3$^\circ$ latitude) of the Equator, it is symmetric about the Equator and extends nearly across the whole length (more than 12000 km) of the Pacific Ocean basin \cite{Iz}. With speeds in excess of 1 m/s, the EUC is one of the fastest permanent currents in the world.

 The flow has nearly two-dimensional character, with small meridional variations. While at depths in excess of about 240 m there is, essentially, an abyssal layer of still water, the ocean dynamics near the surface is quite complex. In this region the wave
motion typically comprises surface gravity waves with amplitudes of 1-2 m and oscillations with an amplitude of 10-20 m at the thermocline (of mean depth between 50 m and 150 m). These
waves interact with the underlying currents.  In that case the velocity is (anti-) parallel to the Earth's angular speed ${\bf \omega},$ so their vector product is zero. This feature distinguishes the dynamics of the equatorial zone from the ocean dynamics at higher latitudes.

The strong stratification confines the wind-driven currents to a shallow near-surface region,  less than 200 m deep. In the Atlantic and Pacific, the westward trade winds induce a westward surface flow
at speeds of 25-75 cm/s, while a jet-like current -- the Equatorial Undercurrent (EUC) -- flows below it toward the East (counter to the surface current), attaining speeds of more than 1 m/s at a depth of nearly 100 m.
The wind-generated equatorial current in the layer above the thermocline is with a strictly monotonic  depth-dependence and exhibits flow-reversal, while beneath the thermocline the current simply decays with increasing depth, being irrelevant in the abyssal region.

 While viscous theory is essential in explaining the generation of the
equatorial current induced by wind forcing, inviscid theory is adequate for the study of non-turbulent wave-current interactions since the relevant Reynolds numbers are very large (see \cite{Mas}).

For some general facts concerning the description of waves interacting with currents we refer to the following reviews and monographs \cite{ACbook, Per, IGJ, TK} and the references therein.  The present study draws from previous single medium irrotational \cite{Zak}, \cite{BenjOlv}, \cite{Milder}, \cite{Miles0}, \cite{Miles} and rotational \cite{Constantin1}, \cite{Constantin2}, \cite{ACbook}, \cite{ConstantinEscher1}, \cite{TelesdaSilva}, \cite{NearlyHamiltonian}, \cite{Wahlen}, \cite{MM} studies as well as from studies of two-media systems such as \cite{BenjBrid}, \cite{BenjBridPart2}, \cite{Craig1}, \cite{Craig2}, \cite{CJ}, \cite{CJ2}, \cite{ConstIvMartin}, \cite{CI}, \cite{Compelli}, \cite{Compelli2},\cite{CompelliIvanov1},  \cite{CompelliIvanov2}, \cite{CHP},\cite{H1},\cite{H2}, \cite{H3}, \cite{M}, \cite{M1}, \cite{M2}. 
 
 The Hamiltonian approach to water waves dynamics has been put forward for the first time by Zakharov \cite{Zak}.
The Hamiltonian formulation describing the two-dimensional nonlinear interaction between coupled surface waves, internal waves, and an underlying current with piecewise constant vorticity, in a two-media fluid overlying a flat bed has been developed in \cite{ConstIvMartin}, \cite{CI}. In the present study we will be following a similar approach, taking into account the shear current structure suggested in \cite{CJ}. Related results for a flat surface (effectively rigid lid) has been studied in \cite{Compelli}
\cite{Compelli2},\cite{CompelliIvanov1},  \cite{CompelliIvanov2}, \cite{CI3}. 

The model equations will be presented in a canonical Hamiltonian form and then small amplitude and long wave approximations will be derived.

\section{Preliminaries}
\noindent{\sc }
The system under study involves two-dimensional surface and internal gravity water waves and a depth dependent current as per Figure 1. 
\begin{figure}
\centering
\includegraphics[width=0.8\textwidth]{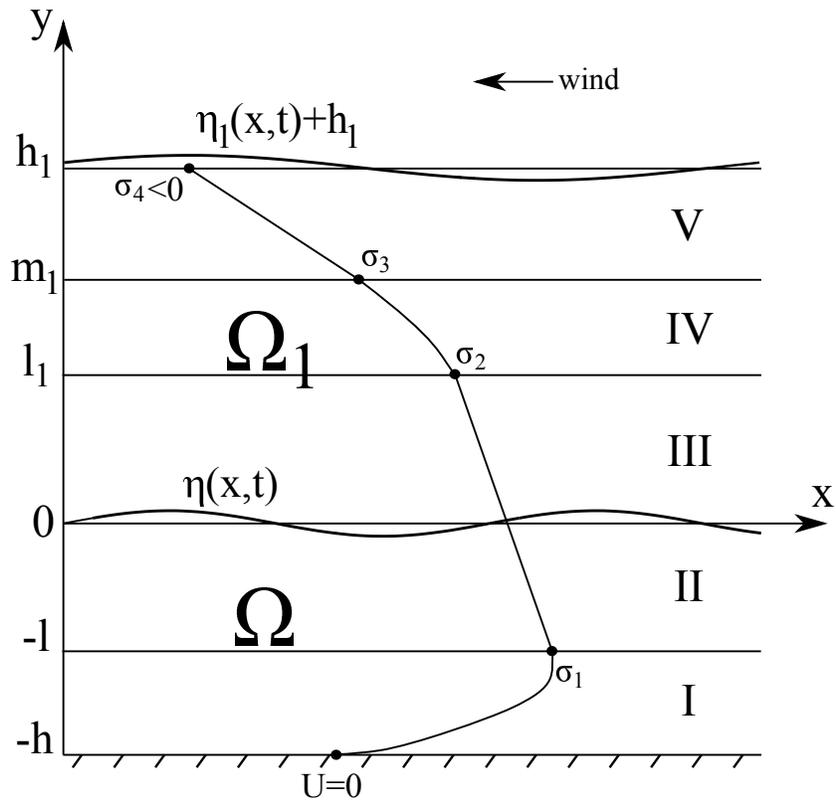}
\caption{System setup. The current profile in layers I and IV is arbitrary as we are only concerned with layers II, III  and V as the internal wave is a free interface between these layers. Continuity of $U(y)$ is assumed in layers I and IV.}
\label{fig:1}       
\end{figure}

The medium underneath the internal wave is defined by the domain $$\Omega(\eta)=\{(x, y)\in\mathbb{R}^2: -h< y < \eta(x,t)\}.$$ This medium is bounded at the bottom by an impermeable flatbed at a depth $-h$. The medium above the internal wave $y=\eta(x,t)$ is the domain $$\Omega_1(\eta,\eta_1)=\{(x, y)\in\mathbb{R}^2: \eta(x,t)< y < h_1+\eta_1(x,t)\}.$$ This medium is regarded as being bounded on top by a surface wave at a hight $y=h_1+\eta_1(x,t)$  moving around the average level $y=h_1.$ Throughout the article the subscript $1$ will be used to mean evaluation for the upper medium $\Omega_1$, and no subscript means evaluation for the lower medium $\Omega$. Subscript $s$ will be used to denote evaluation at the common interface (thermocline/pycnocline), subscript $s_1$ - evaluation at the free surface.

The velocity field is denoted by ${\bf{V}}(x,y,z)=(u,v,0)$ in $\Omega$ and ${\bf{V}}_1(x,y,z)=(u_1,v_1,0)$ in $\Omega_1$. The function $\eta(x,t)$ describes the deviation of the internal wave from its average level $y=0$, i.e. $\ii \eta(x,t) dx=0.$ Similarly, we define the mean of $\eta_1+h_1$ to be the unperturbed surface $y=h_1, $  i.e. $\ii \eta_1(x,t) dx=0.$

A depth dependent current $U(y)$ has the following structure:   
\begin{equation}
\label{U_depth}
  U(y)=      \left\lbrace
        \begin{array}{lcl}
        \sigma_4=\gamma_1 h_1 + \kappa_1<0 \qquad y=h_1,
        \\
        \gamma_1 y + \kappa_1, \qquad m_1\le y \qquad \mbox{ (layer V),}
        \\
        \sigma_3=\gamma_1 m_1 + \kappa_1 \qquad y=m_1,
        \\
        \sigma_2= \gamma l_1+\kappa \qquad  y=l_1, 
        \\
        \gamma y+\kappa,\qquad  -l\le y \le l_1 \qquad \mbox{ (layers II and III),}
        \\
        \sigma_1=-\gamma l+\kappa \qquad y=-l,
        \\
        0 \qquad   y = -h\mbox{ (flatbed),}
        \end{array}
        \right.
\end{equation}
for constants $\sigma_1$, $\sigma_2$, $\sigma_3$, $\sigma_4<0,$ $\kappa$, $\kappa_1,$ $l,$ $l_1$, $\gamma$ and $\gamma_1<0$, where $\kappa$ is the time-independent current velocity at $y=0;$  $\gamma$ and $\gamma_1$ are the non-zero constant vorticity for layers II, III and V, noting that the current is not specified explicitly in layers I and IV, and satisfies only the condition that in the whole fluid body $U(y)$ is a continuous function.
This setup of the wave-current system is motivated in \cite{CJ}, see also \cite{CompelliIvanov1,CompelliIvanov2}.
 The wave motion on the surface is usually confined in the top layer V$^\prime$, defined for depths $y$ such that

$$\text{V}^\prime: \quad m_1 \le y\le  h_1+\eta_1(x,t),$$ and the internal wave is confined in the strip formed of layers II and III:

$$-l\le \eta (x,t)\le l_1. $$ 

\noindent We define also layers II$^\prime$ and III$^\prime$:

\begin{equation}
\begin{split}
\text{II}^\prime: &\quad -l \le y\le  \eta(x,t),\\
\text{III}^\prime: &\quad  \eta(x,t) \le y \le l_1.
\end{split}
\end{equation}

We consider a velocity field, which is defined by the wave-related velocity potentials $\varphi(x,y,t)$ for the domain $\Omega$ and $\varphi_1(x,y,t)$ for the domain $\Omega_1$ as follows:
\begin{equation}
\label{phi_def}
        \left\lbrace
        \begin{array}{lcl}
        u_1  =  \varphi_{1,x} + \gamma_1 y + \kappa_1 \qquad \mbox{ (layer V$^\prime$),} \\
 u_1  = {\varphi}_{1,x} + \gamma y + \kappa \qquad \mbox{ (layer III$^\prime$),}
\\
        v_1  ={\varphi}_{1,y} ,
        \\
         u  =  {\varphi}_{x} + \gamma y + \kappa  \qquad \mbox{ (layer II$^\prime$),} \\
        v  ={\varphi}_{y}.
        \end{array}
        \right.
\end{equation}
We note that this representation separates the wave and current contributions to the velocity in layers II$^\prime$, III$^\prime$ and V$^\prime$, so that the horizontal velocity field in $\Omega$ is nominally separated to a wave and current part, i.e. $u={{\varphi}}_{x}+U(y)$, and similarly $u_1={{\varphi}}_{1,x}+U(y)$ for $\Omega_1,$ see also \cite{CI,ConstIvMartin}.

The respective constant densities $\rho$ and $\rho_1$ of the lower and upper media satisfy the stability condition for immiscibility
\begin{alignat}{2}
\label{stability}
\rho>\rho_1.
\end{alignat}

\noindent The rotationality of the layers II and III is given by constant vorticity \begin{alignat}{2}
\label{rotat}
\gamma=\frac{\sigma_2-\sigma_1}{l+l_1}.
\end{alignat}
The sign of $\gamma$ is not specified and the case  $\gamma =0$ is of course possible.

We assume that for large $|x|$ the amplitude of $\eta$ attenuates and hence make the following assumptions
\begin{alignat}{2}
\label{Assump1}
\lim_{|x|\rightarrow \infty}\eta(x,t)=0, \qquad \lim_{|x|\rightarrow \infty}\eta_1(x,t)=0,
\end{alignat}
\begin{alignat}{2}
\label{Assump1a}
\lim_{|x|\rightarrow \infty}{{\varphi}}(x,y,t)=0, \qquad \lim_{|x|\rightarrow \infty}{{\varphi}}_1(x,y,t)=0.
\end{alignat}
Moreover, we assume 
\begin{alignat}{2}
\label{Assump1b}
-l \le \eta(x,t) < l_1, \qquad  m_1 \le h_1+ \eta_1(x,t)
\end{alignat}

\noindent for all $x$ and $t$. This is an assumption that the surface wave motion takes place only in the \textit{strip} of layer V$^\prime$, and the internal wave motion is only in the \textit{strip} of layers II and III. 

\section{Governing equations}

The governing equations for an inviscid flow are the Euler's equations modified with terms, taking into account the Coriolis force.   

Denoting with $(\bm{u}(x,y,t),\bm{v}(x,y,t))$ the velocity field in $\Omega\cup\Omega_1$, the horizontal component of the velocity field is
\begin{equation}
\bm{u}(x,y,t):=
\left\{\begin{array}{ccc}  u(x,y,t), &{\rm in} & \Omega(\eta),\\ u_1(x,y,t), &{\rm in} &\Omega_1(\eta,\eta_1),\end{array}\right.
\end{equation}
and the vertical component of the velocity field is 
\begin{equation}
\bm{v}(x,y,t):=
\left\{\begin{array}{ccc} v(x,y,t), &{\rm in} & \Omega(\eta),\\ v_1(x,y,t), &{\rm in} &\Omega_1(\eta,\eta_1).\end{array}\right.
\end{equation}
The mass conservation is given by 
\begin{equation}\label{masscons}
\bm{ u}_x+\bm{v}_y=0\,\,\textrm{in}\,\,\Omega\cup\Omega_1.
\end{equation}

The equation of mass conservation \eqref{masscons} ensures the existence of a stream function 
$$\bm{\psi}(t,x,y)=\left\{\begin{array}{ccc}\psi(t,x,y)& {\rm in}&\Omega,\\ \psi_1(t,x,y) & {\rm in} & \Omega_1,\end{array}\right.$$ 
determined up to an additive term that depends only on time, by
\begin{equation}\label{stream_func}
\left\{
\begin{array}{lll}
 u=\psi_y, & v=-\psi_x, & {\rm in} \quad \Omega,\\
u_1=\psi_{1,y}, & v_1=-\psi_{1,x}, & {\rm in}\quad\Omega_1.
\end{array}\right.
\end{equation}

\noindent For convenience we introduce also
\begin{equation}
\bm{\varphi}(x,y,t):=
\left\{\begin{array}{ccc} \varphi (x,y,t), &{\rm in} & \Omega(\eta),\\ \varphi_1(x,y,t), &{\rm in} &\Omega_1(\eta,\eta_1),\end{array}\right.
\end{equation}

\noindent then $$\bm{u}={\bm{\varphi}}_{x}+U(y).$$

Considering equatorial motion the following Coriolis forces per unit mass have to be taken into account:
\begin{equation}
         {\bf{F}}=2\omega \nabla \bm{\psi}=2\omega(-\bm{v}, \bm{u})^T
\end{equation}
with $\omega$ being the rotational speed of the Earth. Then the Euler's equations are 
\begin{equation}
 \left\{\begin{array}{lcl}
        \bm{ u}_t+\bm{u}\bm{u}_x+\bm{v}\bm{u}_y+2\omega \bm{v} &=& -\frac{1}{\bm{\rho}^*}P_x,\\
         \bm{v}_t+\bm{u}\bm{v}_x+\bm{v}\bm{v}_y -2\omega \bm{u} &=& -\frac{1}{\bm{\rho}^*}P_y-g,
        \end{array}\right.
\end{equation}
where $P=P(x,y,t)$ denotes the pressure, $g$ is the gravitational acceleration and $\bm{\rho}^*$ denotes the density of the fluid which, in our case is assumed to be piecewise constant, equal to $\rho$ in $\Omega(\eta)$ and to $\rho_1$ in $\Omega_1(\eta,\eta_1).$

Complementing the equations of motion are the boundary conditions, of which 
\begin{equation}\label{atm}
 P=P_{atm}\,\,{\textrm on}\,\, y=\eta_1 (x)+h_1,
\end{equation}
(with $P_{atm}$ being the constant atmospheric pressure) decouples the motion of the water from that of the air. In addition to \eqref{atm} we have the kinematic boundary conditions
which refer to the flat bed, the interface $y=\eta(x,t)$, the free surface $y=h_1+\eta_1(x,t)$ and reflect the impermeability of these three surfaces. Thus, they read as
\begin{equation}
 \label{kin_fs}
 v_1=\eta_{1,t}+u_1\eta_{1,x}\,\,{\rm on}\,\,y=\eta_1(x,t)+h_1,
\end{equation}
\begin{equation}
 \label{kin_int}
 \begin{array}{lll}
v_1=\eta_t +u_1 \eta_x & {\rm on} & y=\eta(x,t),\\
v=\eta_t+u\eta_x & {\rm on} & y=\eta(x,t),
\end{array}
\end{equation}
and 
\begin{equation}
 v=0 \quad{\rm on}\quad y=-h.
\end{equation}

From the kinematic boundary conditions \eqref{kin_int} one can obtain 

\begin{equation}
 \psi(t,x,\eta(x,t))=\psi_1(t,x,\eta(x,t))=-\int_{-\infty}^x \eta_t(x',t) dx',
\end{equation}

\noindent i.e. $\bm{ \psi}$ is a continuous function across $y=\eta.$ This also implies that the normal velocity field components are continuous across the interface $y=\eta(x,t)$.

We introduce for convenience \begin{equation}\label{chi}
 \chi(x,t)\equiv \psi(t,x,\eta(x,t))=\psi_1(t,x,\eta(x,t)).
\end{equation}
Similarly \begin{equation}\label{chi1}
 \chi_1(x,t)\equiv \psi_1(t,x,\eta_1(x,t)+h_1)=-\int_{-\infty}^x \eta_{1,t}(x',t) dx'.
\end{equation}
With the velocity decompositions  \eqref{phi_def} the kinematic boundary conditions \eqref{kin_fs} and \eqref{kin_int} can now be written as 
\begin{equation}\label{kin_surface}
 \eta_{1,t}=(\varphi_{1,y})_{s_1}-\eta_{1,x} [(\varphi_{1,x})_{s_1}+\gamma_1(h_1+\eta_1)+ \kappa_1]
\end{equation}
and respectively, as
\begin{equation}\label{kin_interface}
 \begin{array}{c}
   \eta_t=( \varphi_{1,y})_{s}-\eta_x [(\varphi_{1,x})_{s}+\gamma\eta+\kappa],\\
 \eta_t=(\varphi_y)_{s}- \eta_x [(\varphi_x)_{s}+\gamma\eta+ \kappa],
  \end{array}
\end{equation}
where the subscript $s_1$ means that we look at traces of the involved functions on the free surface $y=\eta_1(x,t)+h_1$, while the subscript $s$ denotes traces on the interface $y=\eta(x,t)$.
The following notation will be used later on in the paper. Namely, we set
\begin{equation}\label{Phi_notation}
\begin{array}{l}
 \Phi(x,t)=\varphi(x,\eta(x,t),t),\\
 \Phi_1(x,t)=\varphi_1(x,\eta(x,t),t),\\
 \Phi_2(x,t)=\varphi_1(x,h_1+\eta_1(x,t),t).\end{array}
\end{equation}

The Dirichlet-Neumann operator $G(\eta)$ associated to the layer $\Omega(\eta)$ is defined as 
\begin{equation}
 G(\eta)\Phi:=\sqrt{1+\eta_x^2}\frac{\partial\varphi}{\partial n}\Big|_{y=\eta},
\end{equation}
where $n$ denotes the unit outward normal vector to the layer $\Omega(\eta)$ along the interface $y=\eta(x)$.
Recall that $\varphi_1$ is the solution of the boundary value problem
\begin{equation}
 \left\{\begin{array}{ll}
         \Delta\varphi_1 =0 & {\rm in}\quad \Omega(\eta ,\eta_1),\\
         \varphi_1=\Phi_1 & {\rm on}\quad y=\eta,\\
         \varphi_1=\Phi_2 & {\rm on}\quad y=h_1+\eta_1.
        \end{array}\right.
\end{equation}

The Dirichlet-Neumann operator $G_1(\eta,\eta_1)$ associated to the upper media $\Omega_1(\eta,\eta_1)$ is defined through
\begin{equation}
 G_1(\eta,\eta_1)(\Phi_1,\Phi_2):=\begin{pmatrix}-\sqrt{1+\eta_x^2}\frac{\partial \varphi_1}{\partial n}\Big|_{y=\eta}\\
 \sqrt{1+\eta_{1,x}^2}\frac{\partial \varphi_1}{\partial n_1}\Big|_{y=\eta_1 +h_1} \end{pmatrix},
\end{equation}
where, we denote by $n_1$ the unit outward normal vector to $\Omega_1(\eta,\eta_1)$ along the free surface $y=\eta_1(x)+h_1$. 
Of course, $G_1(\eta,\eta_1)$ is a matrix-operator, for which we choose the notation 
\begin{equation}
 G_1(\eta,\eta_1)=\begin{pmatrix}G_{11} & G_{12}\\ G_{21} & G_{22}  \end{pmatrix}.
\end{equation}

Euler's equations can be recast by means of the stream function and of the generalized velocity potential in the given layers in the form of Bernoulli conservation laws as follows: 
$$\varphi_{1,t}+\frac{1}{2}|\nabla\psi_1|^2 -(\gamma_1+2\omega)\psi_1 +\frac{P}{\rho_1}+gy=\tilde{f}_1(t)\quad \text{ in layer V}^\prime $$  where $\tilde{f}_1(t)$ is an arbitrary function of $t$. This is related to the freedom to change $\varphi_1$, if necessary, by a time-dependent factor. Making use of \eqref{atm} we can absorb the constant $P_{atm}$ in the arbitrary function and obtain on $ y=h_1+\eta_1(x,t)$
\begin{equation}\label{Btop}
 \rho_1(\varphi_{1,t})_{s_1}+\frac{\rho_1}{2}|\nabla\psi_1|^2_{s_1}-\rho_1(\gamma_1+2\omega)\chi_1+\rho_1g(h_1+\eta_1)=f_1(t),
\end{equation} for another function 
\begin{equation}\label{f1}
f_1(t)=\rho_1 \tilde{f}_1(t) -P_{atm}.
\end{equation} Similarly, on the interface $y=\eta(x,t)$ the pressure is continuous function and thus it follows that 
\begin{equation}
 \label{Bint}
 \begin{split}
 &\rho\left[(\varphi_t)_s+\frac{|\nabla\psi|_s^2}{2}-(\gamma+2\omega)\chi+g\eta\right]\\
 &=\rho_1\left[(\varphi_{1,t})_s+\frac{|\nabla\psi_1|_s^2}{2}-(\gamma+2\omega)\chi+g\eta\right]+f_2(t),
 \end{split}
\end{equation}

\noindent where $f_2(t)$ is another function of $t$.  The equality \eqref{Bint} can be written as 

\begin{equation}
 \label{Bint1}
 \left[\rho\varphi_t-\rho_1\varphi_{1,t}\right]_s+\frac{\rho|\nabla\psi|_s^2}{2}- \frac{\rho_1|\nabla\psi_1|_s^2}{2}-(\rho-\rho_1)(\gamma+2\omega)\chi+
 (\rho-\rho_1)g\eta=f_2(t).
\end{equation}

\noindent Introducing the constants \begin{equation}  \label{Gamma}
 \begin{split}
 &\Gamma_1=\rho_1(\gamma_1+2\omega),\\
 &\Gamma=(\rho-\rho_1)(\gamma+2\omega),
 \end{split}
\end{equation}

\noindent the variables (suggested as canonical variables in the Hamiltonian formulation in \cite{BenjBrid,BenjBridPart2} )
\begin{equation}
 \label{xi}
 \begin{split}
 &\xi_1(x,t)=\rho_1 \varphi_1(x, \eta_1(x,t)+h_1,t)\equiv \rho_1 \Phi_2(x,t),\\
 &\xi=\rho \varphi(x,\eta(x,t),t) -\rho_1 \varphi_1(x,\eta(x,t),t)\equiv \rho \Phi(x,t) - \rho_1 \Phi_1(x,t)
 \end{split}
\end{equation}
and noting that 

\begin{alignat}{2}
\label{Bern1}
[\rho{{\varphi}_x}-\rho_1{{\varphi}_{1,x}}]_s&= \xi_x-[\rho{{\varphi}_{y}}-\rho_1 {{\varphi}_{1,y}}]_s\eta_x,\\
\label{Bern2}
[\rho{{\varphi}_{t}}-\rho_1{{\varphi}_{1,t}}]_s&= \xi_t-[\rho{{\varphi}_{y}}-\rho_1 {{\varphi}_{1,y}}]_s\eta_t,\\
\label{Bern3}
\rho_1({{\varphi}_{1,t}})_{s_1}&= \xi_{1,t}-\rho_1({{\varphi}_{1,y}})_{s_1}\eta_{1,t},\\
\label{Bern4}
\rho_1({{\varphi}_{1,x}})_{s_1}&= \xi_{1,x}-\rho_1({{\varphi}_{1,y}})_{s_1}\eta_{1,x},
\end{alignat}

\noindent we recast \eqref{Btop} and  \eqref{Bint1} in terms of the velocity potentials and the current parameters, using the relation between the stream function and the velocity potentials that follows from \eqref{stream_func} and \eqref{phi_def}:

\begin{equation}\label{Btop2}
\begin{split}
\xi_{1,t}&+\frac{\rho_1}{2}|\nabla\varphi_1|^2_{s_1}-\rho_1(\varphi_{1,y})_{s_1}[\varphi_{1,y}-\varphi_{1,x}\eta_{1,x}]_{s_1}
+\rho_1g(h_1+\eta_1)\\
&+(\gamma_1 (\eta_1+h_1)+\kappa_1)\xi_{1,x}+\frac{\rho_1}{2}(\gamma_1 (\eta_1+h_1)+\kappa_1)^2-\Gamma_1\chi_1=f_1(t),
\end{split}
\end{equation} 

\begin{equation}
 \label{Bint1a}
 \begin{split}
 \xi_t-&[\rho{{\varphi}_{y}}-\rho_1 {{\varphi}_{1,y}}]_s[{\varphi}_{y}-{\varphi}_{x} \eta_x]_s+\frac{\rho|\nabla\varphi|_s^2}{2}- \frac{\rho_1|\nabla\varphi_1|_s^2}{2}+
 (\rho-\rho_1)g\eta\\
 &+(\gamma \eta+\kappa)\xi_{x}+\frac{\rho-\rho_1}{2}(\gamma \eta+\kappa)^2-\Gamma\chi=f_2(t).
 \end{split}
\end{equation}

The balance of all quantities when $x\to \pm \infty$ gives

\begin{equation} \label{f12}
f_1=\rho_1 \left(g h_1+\frac{1}{2}(\gamma_1h_1+\kappa_1)^2\right), \qquad f_2=\frac{1}{2}(\rho-\rho_1)\kappa^2.
\end{equation}

Recall that in the absence of current and Coriolis force the system is Hamiltonian and can be represented in the form 
\begin{equation}\label{Heta}
\begin{split}
&\xi_{1,t}=-\delta_{\eta_1}H_0, \qquad \xi_{t}=-\delta_{\eta}H_0, \\
&\eta_{1,t}=\delta_{\xi_1}H_0, \qquad \eta_{t}=\delta_{\xi}H_0,
\end{split}
\end{equation} 

\noindent where $H_0(\eta, \eta_1,\xi,\xi_1)$ is the corresponding Hamiltonian. $H_0$ is evaluated in terms of the canonical variables, see \cite{Craig1,Craig2} for details:

\begin{align}
H_0(\eta, \eta_1,\xi,\xi_1)&=\frac{1}{2}\int_{\mathbb{R}}\begin{pmatrix}\xi\\ \xi_1\end{pmatrix}^{t}
\begin{pmatrix}G_{11}B^{-1}G(\eta) & -G(\eta)B^{-1}G_{12}\\-G_{21}B^{-1}G(\eta) & -\frac{\rho}{\rho_1}G_{21}B^{-1}G_{12} +\frac{1}{\rho_1}G_{22}\end{pmatrix}\begin{pmatrix}\xi\\ \xi_1\end{pmatrix}\,dx\nonumber\\
&+\frac{1}{2}\int_{\mathbb{R}}\left( g(\rho-\rho_1)\eta^2 + g\rho_1 \eta_1^2 +2g\rho_1 h_1 \eta_1    \right) \, dx ,                   \nonumber  \label{H0}
\end{align}

\noindent where the operator $B\equiv \rho_1 G(\eta)+\rho G_{11}.$

 Hence we can represent all terms not related to vorticity as variational derivatives of $H_0:$ 
\begin{equation}\label{Hxi}
\begin{split}
\xi_{1,t}&+\delta_{\eta_{1}}H_0+(\gamma_1( \eta_1+h_1)+\kappa_1)\xi_{1,x}+\frac{\rho_1}{2}\left(\gamma_1 (\eta_1+h_1)+\kappa_1\right)^2-\Gamma_1\chi_1=f_1(t),\\
\xi_t& +\delta_{\eta}H_0+   (\gamma \eta+\kappa)\xi_{x}+\frac{\rho-\rho_1}{2}(\gamma \eta+\kappa)^2-\Gamma\chi=f_2(t).
\end{split}
\end{equation} 

\noindent In addition, \eqref{kin_surface}, \eqref{kin_interface} can be written as

\begin{equation}\label{Heta}
\begin{split}
 \eta_{1,t}&=\delta_{\xi_1}H_0 -[\gamma_1(h_1+\eta_1)+ \kappa_1]\eta_{1,x},\\
 \eta_t &=\delta_{\xi}H_0 -(\gamma\eta+\kappa)\eta_x.\\
   \end{split}
\end{equation}

\noindent Now we are in a position to write the equations \eqref{Hxi}, \eqref{Heta} in the form

\begin{equation}\label{Hxieta}
\begin{split}
\xi_{1,t}&=-\delta_{\eta_{1}}H + \Gamma_1 \chi_1, \qquad \xi_t =-\delta_{\eta}H+\Gamma\chi, \\
 \eta_{1,t}&=\delta_{\xi_1}H, \qquad  \qquad \eta_t =\delta_{\xi}H, 
\end{split}
\end{equation} 

\noindent where
\begin{equation}\label{Hnew}
\begin{split}
H&(\eta, \eta_1, \xi, \xi_1)=H_0-\kappa\ii \xi \eta_x dx-\ii  \gamma\eta\eta_x \xi dx
+\frac{\rho-\rho_1}{6\gamma}\ii (\gamma \eta+\kappa)^3dx \\
&-\kappa_1\ii \xi_1 \eta_{1,x} dx-\ii  \gamma_1(\eta_1+h_1)\eta_{1,x} \xi_1 dx +\frac{\rho_1}{6\gamma_1}\ii (\gamma_1 (\eta_1+h_1)+\kappa_1)^3dx \\
&-f_1(t) \ii \eta_1 dx - f_2(t) \ii \eta dx -H_{00}.
\end{split}
\end{equation} 

$$H_{00}=\frac{\rho-\rho_1}{6\gamma}\ii \kappa^3dx +\frac{\rho_1}{6\gamma_1}\ii (\gamma_1 h_1+\kappa_1)^3dx $$ is an integral with a constant Hamiltonian density and zero variational derivatives which keeps the overall Hamiltonian density decaying to zero at $x\to \pm \infty.$ The \textit{ghost} terms with $f_{1,2}$ are un-physical (i.e. their values do not affect the measurable physical quantities such as velocities and elevations), since by definition $$\ii \eta dx=0, \qquad \ii \eta_1 dx=0. $$  Nevertheless their variational derivatives produce the $f_{1,2}$ terms in \eqref{Hxi} which are also un-physical and can in principle be absorbed in the definition of the potentials $\xi,$ $\xi_1,$  in our case, in such a way that $\xi,$ $\xi_1$  tend to zero when $x\to \pm \infty.$ From \eqref{f12} it follows that $H$ does not contain terms, linear in the field variables. The lowest order terms are quadratic in the field variables and they produce the linearised equations. 

The expression for the Hamiltonian \eqref{Hnew} can be obtained alternatively by evaluation of the total energy of the fluid (up to a constant Hamiltonian density):
$$H=\int\int_{\Omega\cup\Omega_1}\bm{\rho}^*\left\{\frac{\bm{u}^2+\bm{v}^2}{2}+gy\right\}dydx.$$
Taking into account the stratification of the fluid the above expression can be rewritten as
\begin{equation}
\begin{split}
H=&\frac{1}{2}\int_{\mathbb{R}}\int_{-h}^{\eta(x,t)}\rho(u^2+v^2)dydx^{\prime}+\frac{1}{2}\int_{\mathbb{R}}\int_{\eta(x,t)}^{h_1+\eta_1(x,t)}\rho_1(u_1^2+v_1^2)dydx^{\prime} \\
&+\int_{\mathbb{R}}\int_{-h}^{\eta(x,t)}g\rho y\, dy dx^{\prime}+\int_{\mathbb{R}}\int_{\eta(x,t)}^{h_1+\eta_1(x,t)} g\rho_1 y\,dydx^{\prime}. 
\end{split}
\end{equation}

The computations follow the routine from \cite{ConstIvMartin, CompelliIvanov1,CompelliIvanov2}. We mention only that due to the two-dimensional character of the dynamics, the final expression \eqref{Hnew} depends only on the variables on the surface and on the interface. It is an important feature, that there is no contribution from layers I and IV where the wave motion on $s$ and $s_1$ does not take place \cite{CompelliIvanov1,CompelliIvanov2}. The terms related to the internal wave are the same as in \cite{CompelliIvanov2} where flat surface approximation is considered.

\section{Hamiltonian dynamics}

The ideas for the Hamiltonian formulation of water waves coupled to a flow with a constant vorticity originate from \cite{NearlyHamiltonian} followed by \cite{Wahlen}. For internal waves with vorticity the problem is studied in  \cite{ConstIvMartin, CI, Compelli,Compelli2}.  

The evolution equations \eqref{Hxieta} can be written in a canonical Hamiltonian form in terms of the variables 
\begin{equation}\label{change_of_variables}
 \begin{array}{l}
  z=\xi+\frac{\Gamma}{2}\int_{-\infty}^{x}\eta(x^{\prime},t)\,dx^{\prime},\\
  z_1=\xi_1+\frac{\Gamma_1}{2}\int_{-\infty}^{x}\eta_1(x^{\prime},t)\,dx^{\prime}.
 \end{array}
\end{equation}

The system described by the phase space variables $\eta,\eta_1, z,z_1$ is Hamiltonian. More precisely, 
$$\frac{\delta H}{\delta \eta}=-z_t,\qquad \frac{\delta H}{\delta z}=\eta_t,$$
$$\frac{\delta H}{\delta \eta_1}=-z_{1,t},\qquad \frac{\delta H}{\delta z_1}=\eta_{1,t}.$$
The proof follows the lines of the one in \cite{ConstIvMartin}. Since 

\begin{equation}
\begin{split}
 \chi(x,t)& =-\int_{-\infty}^x \eta_t(x',t) dx'=\int_{-\infty}^x \frac{\delta H}{\delta \eta(x',t)}dx', \\
\chi_1(x,t)& =-\int_{-\infty}^x \eta_{1,t}(x',t) dx'=\int_{-\infty}^x \frac{\delta H}{\delta \eta_1(x',t)}dx',
\end{split} 
\end{equation}

\noindent the system of equations \eqref{Hxieta} can be written as 

\begin{equation} \label{Hamilt1}
\begin{split}
\dot{\xi_k}=&-\frac{\delta H}{\delta \eta_k}-\Gamma_k \int_{-\infty}^{x}\frac{\delta H}{\delta \xi_k(x')}dx',   \\
\dot{\eta_k}=&\frac{\delta H}{\delta \xi_k},
\end{split}
\end{equation}

\noindent where for convenience $k=0,1$, $\eta_0\equiv \eta$, $\xi_0\equiv \xi,$ $\chi_0\equiv \chi$ and $\Gamma_0\equiv \Gamma.$  This is an equivalent Hamiltonian form, in terms of the original variables and with respect to the Poisson bracket (PB)

\begin{equation}\label{PBdef}
\begin{split}
\{A,B\}=&\sum_{k=0}^{1}\int_{\mathbb{R}} \left(\frac{\delta A}{\delta \eta_k (x)}\frac{\delta B}{\delta \xi_k (x)}-\frac{\delta A}{\delta \xi_k (x)}\frac{\delta B}{\delta \eta_k (x)} \right) dx \\
&- \Gamma_k\int_{\mathbb{R}} \left(\frac{\delta A}{\delta \xi_k (x)}\int_{-\infty}^{x}\frac{\delta B}{\delta \xi_k (x')}dx'\right)dx
\end{split}
\end{equation}

\noindent i.e. \begin{equation} \label{Hamilt2}
\begin{split}
\dot{\xi_k}=&\{\xi_k, H\} \qquad k=0,1, \\
\dot{\eta_k}=&\{\eta_k,H\} \qquad k=0,1.
\end{split}
\end{equation}

Here we have to specify the spaces where the functionals $A,B$ may belong. The antisymmetry of the PB requires for example 

\begin{equation}\label{antisymetry}
\int_{\mathbb{R}} \left(\frac{\delta A}{\delta \xi_k (x)}\int_{-\infty}^{x}\frac{\delta B}{\delta \xi_k (x')}dx'\right)dx=-\int_{\mathbb{R}} \left(\frac{\delta B}{\delta \xi_k (x)}\int_{-\infty}^{x}\frac{\delta A}{\delta \xi_k (x')}dx'\right)dx
\end{equation}

\noindent which is possible (e.g. due to integration by parts) if and only if

 \begin{equation}
\int_{\mathbb{R}} \frac{\delta A}{\delta \xi_k (x)}dx \int_{\mathbb{R}}\frac{\delta B}{\delta \xi_k (x')}dx'=0.
\end{equation}

\noindent Thus, at least one of the functionals in the PB should satisfy
 \begin{equation}
\int_{\mathbb{R}} \frac{\delta A}{\delta \xi_k (x)}dx =0.
\end{equation}

\noindent Since in  (\ref{Hamilt2}) one of the functionals is always $H$ and $$\int_{\mathbb{R}} \frac{\delta H}{\delta \xi_k (x)}dx =\int_{\mathbb{R}}\partial_t \eta_k(x,t)=\lim_{x\to \infty}\chi_k(x,t)=0 .$$

\section{The pressure in the  body of the fluid}

The pressure in the body of the fluid can be evaluated from the functions  $\bm{\varphi}$ and $\bm{\psi}.$   They can be recovered from $\xi$ and $\xi_1(x,t).$  In addition, there is an interdependency between $\bm{\varphi}$ and $\bm{\psi}.$ since $$\bm{\varphi}+i\left(\bm{\psi}- \int_{-h}^{y } U(y') d y' \right )$$ is an analytic function of the variable $z=x+i y$  in the domain $\Omega_1\cup \Omega.$ Thus, the corresponding analytic functions in $\Omega$ and $\Omega_1$ can be recovered from their values at the boundaries of $\Omega$ and $\Omega_1,$ i.e. from  $\Phi(x,t)$, $\Phi_1(x,t)$ and $\Phi_2(x,t).$

 From the definition of the Dirichlet-Neumann operators and \eqref{kin_interface} we have that
 $$G_{11}\Phi_1 +G_{12}\Phi_2=\varphi_{1,x}\eta_x-\varphi_{1,y}=-(\eta_t+(\gamma\eta+\kappa)\eta_x),$$
 and 
 $$G(\eta)\Phi=-\varphi_x\eta_x+\varphi_y= \eta_t+(\gamma\eta+\kappa)\eta_x.$$
 Adding up the previous two relations we obtain
 \begin{equation}\label{G-G_1}
  G_{11}\Phi_1 +G_{12}\Phi_2 +G\Phi=0.
 \end{equation}
Using \eqref{G-G_1}, and $B:=\rho_1 G(\eta)+\rho G_{11}$ and recalling that $$\xi=\rho\Phi-\rho_1\Phi_1,\quad \xi_1=\rho_1\Phi_2$$ we can write $\Phi,\Phi_1,\Phi_2$ in terms of the Hamiltonian variables $\xi$ and $\xi_1$ as 
follows

 \begin{align}
 \Phi&= B^{-1}\left(G_{11}\xi -G_{12}\xi_1\right)\label{Phi},\\
  \Phi_1&= B^{-1}\left(-G(\eta)\xi-\frac{\rho}{\rho_1}G_{12}\xi_1\right)\label{Phi1},\\
  \Phi_2 &=\frac{1}{\rho_1}\xi_1\label{Phi2}.
 \end{align}

In the layers with a fixed constant vorticity $\bm{\gamma}$ we have Bernoulli conservation laws  
\begin{equation}\label{Ber_gen1}
\bm{\varphi}_t+\frac{1}{2}|\nabla\bm{\psi}|^2+\frac{P}{\rho^*}-(\bm{\gamma}+2\omega)\bm{\psi}+g y=\tilde{\bm{f}}(t).
\end{equation}

\noindent From \eqref{f1} and \eqref{f12} for the layer V$^\prime$ we have 

\begin{equation}\label{tdf1} \tilde{\bm{f}} =\tilde{f}_1= \frac{f_1+P_{atm}}{\rho_1}=\frac{P_{atm}}{\rho_1}+gh_1
+\frac{1}{2}(\gamma_1h_1+\kappa_1)^2  \end{equation} 

\noindent Another possible derivation of \eqref{tdf1} is the following one. We can evaluate $\tilde{f}_1$ using the asymptotic values in \eqref{Ber_gen1} at $x\to \pm \infty$ where no wave motion takes place ($y=h_1$) and thus $\psi_{1,x} \to 0$ and $\psi_{1,y} \to \gamma_1 h_1 + \kappa_1$ are given by the velocity of the current. Moreover, $\psi_1$ on the surface is zero when $x\to \pm \infty$ due to \eqref{chi1}. Hence \begin{equation}\label{P5}
\begin{split}
P(x,y,t)=& P_{atm}- \rho_1\left({\varphi_1}_t+\frac{1}{2}|\nabla{\psi_1}|^2-({\gamma_1}+2\omega){\psi_1}
+g (y-h_1) \right) \\
+&\frac{\rho_1}{2 }(\gamma_1h_1+\kappa_1)^2 \qquad \text{for} \quad m_1\le y\le h_1+\eta_1(x,t). 
\end{split}
\end{equation}

We can apply the same approach in deriving the pressure in layers III$^{\prime}$ and II$^{\prime}$.  At $y=0$ and $x\to \pm \infty$ due to the lack of a wave motion the pressure is $P_0=P_{atm}+\rho_1 g h_1. $  For the layer III$^\prime$ we have

\begin{equation}\label{P3}
\begin{split}
P(x,y,t)=& P_{atm}- \rho_1\left({\varphi_1}_t+\frac{1}{2}|\nabla{\psi_1}|^2-({\gamma}+2\omega){\psi_1}
+g (y-h_1) \right) +\frac{\rho_1}{2}\kappa^2 \\
&\quad  \text{for} \quad \eta(x,t)\le y\le l_1  ,
\end{split}
\end{equation}

\noindent For the layer II$^\prime ,$  \begin{equation}\label{P2}
\begin{split}
P(x,y,t)=& P_{atm} +\rho_1 gh_1 -\rho\left({\varphi}_t+\frac{1}{2}|\nabla{\psi}|^2-({\gamma}+2\omega){\psi}
+g y \right)+\frac{\rho}{2}\kappa^2\\
& \quad \text{for} \quad   -l\le y \le \eta(x,t) .
 \end{split}
\end{equation}

\section{Scales}

Let us introduce non-dimensional variables (without bars) related to the dimensional (barred) as follows:

\begin{equation} \label{nondim}
\begin{split}
\bar{t}=&\frac{h_1}{\sqrt{gh_1}}t, \qquad \bar{x}=h_1x, \qquad \bar{y}=h_1y, \qquad \bar{\eta}=a\eta, \qquad \bar{\eta}_1= a\eta_1, \\
 \bar{u}=&\sqrt{gh_1}u, \qquad  \bar{u}_1=\sqrt{gh_1}u_1,\qquad \bar{v}=\sqrt{gh_1} v, \qquad \bar{\kappa}=\sqrt{gh_1}\kappa, \qquad
\\
\bar{\kappa}_1=&\sqrt{gh_1}\kappa_1, \qquad  \bar{\gamma}=\frac{\sqrt{gh_1}}{h_1}\gamma, \qquad  \bar{\gamma}_1=\frac{\sqrt{gh_1}}{h_1}\gamma_1,  \qquad \varepsilon=\frac{a}{h_1}.
\end{split}
\end{equation}

The constant $a$ represents the average amplitude of the waves under consideration, $\varepsilon$ is a small parameter which will be used to separate the order of the terms in the model.

From $\bar{v}=\bar{\eta}_{\bar{t}}+\bar{u}\bar{\eta}_{\bar{x}}$ it follows

\begin{equation} \label{nondim3}
v=\varepsilon(\eta_t+u\eta_x),
\end{equation} therefore, if $\mathcal{O}(\eta_t)=1$ then $\mathcal{O}(v)=\varepsilon$ and thus the dimensional expression with $\mathcal{O}(v)=1$ (and similar for $v_1$) should be 

\begin{equation} \label{nondim4}
\bar{v}=\varepsilon \sqrt{gh_1} v, \qquad \bar{v_1}=\varepsilon \sqrt{gh_1} v_1.
\end{equation}

Since $v$ is a $y$-derivative of the velocity potential, and with the adopted definitions $\bar{\varphi}=\varepsilon h_1 \sqrt{gh_1}\varphi $ etc. and thus

\begin{equation} \label{nondim5}
\bar{\xi}=\varepsilon \rho h_1 \sqrt{gh_1} \xi, \qquad \bar{\xi}_1=\varepsilon \rho h_1 \sqrt{gh_1} \xi_1.
\end{equation}  The scales for $u$, $u_1$ etc. do not change - their dominant parts are the vorticity  and current components of order 1; only the 'wave' component (which is $x$ derivative of  $\varphi $) is of order $\varepsilon$.
The Dirichlet-Neumann operators have the following structure (e.g. any of the introduced operators $G$, $G_{ij}$):

\begin{equation} \label{DN}
\bar{G}=\bar{G}^{(0)} +\bar{ G}^{(1)}+\bar{G}^{(2)}+ \ldots
\end{equation} 
where $\bar{G}^{(n)}\sim \bar{\eta}^n \partial_{\bar{x}}^{n+1}$ i.e. $\bar{G}^{(n)}=\frac{\varepsilon^n}{h_1}G^{(n)}$:  

\begin{equation} \label{DN}
\bar{G}=\frac{1}{h_1}\left(G^{(0)} +\varepsilon G^{(1)}+\varepsilon^2 G^{(2)}+ \ldots \right).
\end{equation} 

With this scaling and ignoring the linear terms, whose average is 0, the Hamiltonian can be expanded as 
\begin{equation} \label{Hexp}
\bar{H}= \rho g h_1^3\left( \varepsilon^2 H^{(2)} +\varepsilon^3 H^{(3)}+ \ldots \right)
\end{equation}

\noindent where

\begin{align}
H^{(2)}&=\frac{1}{2}\int_{\mathbb{R}}\begin{pmatrix}\xi\\ \xi_1\end{pmatrix}^{t}
\begin{pmatrix}G_{11}B^{-1}G(\eta) & -G(\eta)B^{-1}G_{12}\\-G_{21}B^{-1}G(\eta) & -\frac{\rho}{\rho_1}G_{21}B^{-1}G_{12} +\frac{1}{\rho_1}G_{22}\end{pmatrix}^{(0)}\begin{pmatrix}\xi\\ \xi_1\end{pmatrix}\,dx\nonumber\\
&+\frac{1}{2} (\rho-\rho_1)(g+\gamma \kappa)\int_{\mathbb{R}}\eta^2 dx + \frac{1}{2}\rho_1 (g+\gamma_1^2 h_1 +\gamma_1 \kappa_1) \int_{\mathbb{R}}\eta_1^2   \, dx   \nonumber \\
  &   - \kappa \int_{\mathbb{R}} \xi \eta_x \, dx  - (\kappa_{1} +\gamma_1 h_1)\int_{\mathbb{R}} \xi_1 \eta_{1,x} \, dx 
                   \label{H2}
\end{align}

\noindent where the leading order, $\mathcal{O}(1)$ expression for the operators is (does not depend on $\eta$, $\eta_1 $) 

\begin{equation}
\begin{pmatrix}G_{11} & G_{12}\\G_{21} & G_{22}\end{pmatrix}^{(0)}=\begin{pmatrix} D\coth(h_1D) & -D\text{csch}(h_1D)\\-D\text{csch}(h_1D) & D\coth(h_1D) \end{pmatrix},
\end{equation}

\noindent where $D=-i\partial_x$. The quadratic part produces the linearised equations. Similarly

\begin{align}
H^{(3)}&=\frac{1}{2}\int_{\mathbb{R}}\begin{pmatrix}\xi\\ \xi_1\end{pmatrix}^{t}
\begin{pmatrix}G_{11}B^{-1}G(\eta) & -G(\eta)B^{-1}G_{12}\\-G_{21}B^{-1}G(\eta) & -\frac{\rho}{\rho_1}G_{21}B^{-1}G_{12} +\frac{1}{\rho_1}G_{22}\end{pmatrix}^{(1)}\begin{pmatrix}\xi\\ \xi_1\end{pmatrix}\,dx\nonumber\\
& -\gamma \int_{\mathbb{R}} \xi \eta \eta_x \, dx - \gamma_1 \int_{\mathbb{R}} \xi_1 \eta_1 \eta_{1,x} \, dx +\frac{1}{2}\int_{\mathbb{R}}\left( (\rho-\rho_1) \gamma^2\frac{\eta^3}{3} +\rho_1 \gamma_1^2\frac{\eta_1^3}{3} \right) dx.         
 \label{H3}\end{align}
 
 \noindent The order $\varepsilon$ terms $G^{(1)}$ of the operators are given in \cite{Craig1,Craig2}.

\section{Linearised equations}

The Hamiltonian equations with a Hamiltonian $H^{(2)}$ are the linearised equations:

\begin{equation} \label{newlinsys2}
\begin{split}
\xi_t=&-\kappa\xi_x-(\rho-\rho_1)(g+ \gamma \kappa)\eta-\Gamma \partial^{-1}\eta_t,    \\
\eta_t=&-\kappa \eta_x+ \frac{D\tanh(hD)\coth(h_1D)}{\rho \coth(h_1D)+\rho_1 \tanh(hD)}\xi + \frac{D\tanh(hD)\text{csch}(h_1D)}{\rho \coth(h_1D)+\rho_1 \tanh(hD)}\xi_1 , \\
\xi_{1,t}=&-(\gamma_1h_1+\kappa_1)\xi_{1,x}-\rho_1(g+\gamma_1^2h_1+\gamma_1 \kappa_1)\eta_1-\Gamma_1 \partial^{-1}\eta_{1,t} , \\
\eta_{1,t}=&-(\gamma_1h_1+\kappa_1)\eta_{1,x}+ \frac{D\tanh(hD)\text{csch}(h_1D)}{\rho \coth(h_1D)+\rho_1 \tanh(hD)}\xi \\
&\qquad + \frac{D\big(\tanh(hD)\coth(h_1D)+\frac{\rho}{\rho_1}\big)}{\rho \coth(h_1D)+\rho_1 \tanh(hD)}\xi_1,
\end{split}
\end{equation}
 
 \noindent where $\partial^{-1}$ is the inverse of $\partial_x$. We can change the coordinates via a linear transformation according to $\partial_T=\partial_t +\kappa \partial_x$:

\begin{equation} \label{newlinsys3}
\begin{split}
\xi_T=&-(\rho-\rho_1)(g-2\omega \kappa )\eta- (\rho-\rho_1)(\gamma+2\omega) \partial^{-1}\eta_T, \\ 
\eta_T=& \frac{D\tanh(hD)\coth(h_1D)}{\rho \coth(h_1D)+\rho_1 \tanh(hD)}\xi + \frac{D\tanh(hD)\text{csch}(h_1D)}{\rho \coth(h_1D)+\rho_1 \tanh(hD)}\xi_1, \\
\xi_{1,T}=&-(\gamma_1h_1+\kappa_1 - \kappa)\xi_{1,x}-\rho_1[g+\gamma_1^2h_1+\gamma_1(\kappa_1 - \kappa)-2\omega \kappa]\eta_1 \\ 
&-\rho_1(\gamma_1 + 2 \omega) \partial^{-1}\eta_{1,T} ,\\
\eta_{1,T}=&-(\gamma_1h_1+\kappa_1-\kappa)\eta_{1,x}+ \frac{D\tanh(hD)\text{csch}(h_1D)}{\rho \coth(h_1D)+\rho_1 \tanh(hD)}\xi \\
&+ \frac{D\big(\tanh(hD)\coth(h_1D)+\frac{\rho}{\rho_1}\big)}{\rho \coth(h_1D)+\rho_1 \tanh(hD)}\xi_1.
\end{split}
\end{equation}

Note that the equations contain Coriolis terms dependent on $\omega.$ Usually $\kappa$ is of magnitude several m/s, $2\omega=1.46\times 10^{-4}$ s$^{-1}$ and hence $2\omega\kappa \ll g:$

\begin{equation} \label{newlinsys4}
\begin{split}
\xi_T=&-(\rho-\rho_1)g\eta- (\rho-\rho_1)(\gamma+2\omega) \partial^{-1}\eta_T, \\ 
\eta_T=& \frac{D\tanh(hD)\coth(h_1D)}{\rho \coth(h_1D)+\rho_1 \tanh(hD)}\xi + \frac{D\tanh(hD)\text{csch}(h_1D)}{\rho \coth(h_1D)+\rho_1 \tanh(hD)}\xi_1, \\
\xi_{1,T}=&-a\xi_{1,x}-\rho_1(g+a\gamma_1)\eta_1-\rho_1(\gamma_1 + 2 \omega) \partial^{-1}\eta_{1,T}, \\
\eta_{1,T}=&-a\eta_{1,x}+ \frac{D\tanh(hD)\text{csch}(h_1D)}{\rho \coth(h_1D)+\rho_1 \tanh(hD)}\xi \\
&+ \frac{D\big(\tanh(hD)\coth(h_1D)+\frac{\rho}{\rho_1}\big)}{\rho \coth(h_1D)+\rho_1 \tanh(hD)}\xi_1,
\end{split}
\end{equation}

\noindent where we introduced the notation $$a=\gamma_1h_1+\kappa_1 - \kappa.$$ Next we search for solutions, proportional to 

 \begin{equation}
\begin{split}
e^{i(kx-\Omega_0(k)T)} , 
\end{split}
\end{equation}

\noindent where $\Omega_0(k)$ is the dispersion law for the wave, the wave speed is $$ c_0(k)=\frac{\Omega_0(k)}{k}.$$ Such a solution would be an eigenfunction for $D$ with a corresponding eigenvalue $k$. From the system (\ref{newlinsys4}) one can express $\xi$, $\xi_1$ as

\begin{equation} \label{etas}
\begin{split}
\xi=&\frac{i(\rho-\rho_1)}{k}\left(\gamma + 2 \omega -\frac{g}{c_0(k)}\right)\eta,  \\
\xi_1=&\frac{i\rho_1}{k} \left(\gamma_1+2\omega +\frac{2a\omega -g}{c_0(k)-a}\right)\eta_1. \end{split}
\end{equation}

The remaining two equations lead to a linear homogeneous system for $\eta$ and $\eta_1$. We introduce the notations

\begin{equation} \label{definitions}
\begin{split}
\mu(k)=& \frac{\rho \tanh(hk)\coth(h_1k)}{\rho \coth(h_1k)+\rho_1 \tanh(hk)}, \\
f(k)=& \frac{\rho \tanh(hk)\text{csch}(h_1k)}{\rho \coth(h_1k)+\rho_1 \tanh(hk)},\\
\theta(k)=&\frac{\rho\big(\tanh(hk)\coth(h_1k)+\frac{\rho}{\rho_1}\big)}{\rho \coth(h_1k)+\rho_1 \tanh(hk)}.
\end{split}
\end{equation}

The second equation of (\ref{newlinsys4}) gives  
\begin{equation} \label{c1eq1}
\left[c_0+ \frac{(\rho\!-\!\rho_1)\mu(k)}{\rho k }\left(\!\gamma\!+\! 2 \omega \!-\!\frac{g}{c_0  }\right)\right] \eta  + \frac{\rho_1 f(k)}{\rho k}\left(\!\gamma_1\!+\! 2 \omega\! +\!\frac{2a\omega -g}{c_0-a}\right)\eta_1 =0. \\ 
\end{equation}

Then the last equation of (\ref{newlinsys4}) leads to

\begin{equation} \label{c1eq}
 \frac{(\rho-\rho_1)f(k)}{\rho k}\left(\!\gamma\!+\! 2 \omega\! -\!\frac{g}{c_0  }\right) \eta  + \left[\! c_0\!-\!a\!+\!\frac{ \rho_1 \theta(k)}{\rho k}\left(\!\gamma_1\!+ \!2 \omega\! +\!\frac{2a\omega -g}{c_0-a}\right)\right]\eta_1 =0.\\ 
 \end{equation}

The compatibility of the two equations gives a 4-th order equation for $c_0(k)$:

\begin{equation} \label{ck}
\begin{split}
&\left[c_0+ \frac{\rho-\rho_1}{\rho }\frac{\mu(k)}{ k}\left(\gamma+ 2 \omega -\frac{g}{c_0  }\right)\right]\left[ c_0-a+\frac{\rho_1}{\rho}\frac{ \theta(k)}{k}\left(\gamma_1+ 2 \omega +\frac{2a\omega -g}{c_0-a}\right)\right]
\\
&=  \frac{\rho_1(\rho-\rho_1)}{\rho^2 }\frac{f^2(k)}{ k^2}\left(\gamma+ 2 \omega -\frac{g}{c_0  }\right) \left(\gamma_1+ 2 \omega +\frac{2a\omega -g}{c_0-a}\right).
\end{split}
\end{equation}

The last formula generalises the irrotational one from \cite{Craig2}. In addition, from \eqref{c1eq}, \eqref{c1eq1} one can determine if $\eta$ and $\eta_1$ have the same or an opposite polarity for each possible propagation speed $c_0$. The velocity $c_0(k)$ is relative to an observer moving together with the flow at $y=0$, i.e. with velocity $\kappa.$ The wave speed for a stationary observer therefore is $$c(k)=c_0(k)+\kappa .$$ 

\section{Long waves approximation}

In the long waves approximation the physical scales are measured by the dimensionless parameter like $\delta=\frac{h_1}{\bar{L}}$. We will study the equations under the additional approximation that the wavelengths $L$ are much bigger than $h$ and $h_1$. Since $$\bar{L}=h_1 L \Rightarrow \frac{1}{L}=\frac{h_1}{\bar{L}}=\delta. $$ Thus the wave number $k=2\pi/L=2\pi \delta$ and  $\mathcal{O}(k)=\delta $. We further assume that $\delta^2=\mathcal{O}(\varepsilon).$
Recall that the operator $D$ has an eigenvalue $k$, thus we shall keep in mind that $\mathcal{O}(D)=\delta. $ Moreover the $x$-derivative of the velocity potentials do not get an extra factor of $\delta$ since $\mathcal{O}(\bar{v})=\varepsilon$ remains unchanged and similarly $v_1$. In other words the 'wave' component of $u$ let's call it $\tilde{u}=\varphi_x$ is of order $\varepsilon$ and similarly $\tilde{u}_1=\varphi_{1,x}$ is of order $\varepsilon$. Despite the assumption  $\delta^2=\mathcal{O}(\varepsilon)$ we will keep both scales $\delta$ and $\varepsilon$ in order to keep track of the origin of the various terms. We will keep track only of the scale variables $\varepsilon, \delta$ and not of the other dimensional factors. For example, $H^{(2)}$ itself contains the following type of terms:

\begin{align}
H^{(2)}&=\varepsilon^2\frac{1}{2}\int_{\mathbb{R}}\left(\frac{h}{\rho}\tilde{u}^2
+ \frac{2h}{\rho}\tilde{u}\tilde{u}_1 + \left(\frac{h}{\rho} +\frac{h_1}{\rho_1}\right )\tilde{u}_1^2  \right) \, dx \nonumber \\
&=\varepsilon^2\frac{1}{2}(\rho-\rho_1)(g+\gamma \kappa)\int_{\mathbb{R}}\eta^2 \, dx +\varepsilon^2\frac{1}{2}\rho_1(g+\gamma_1^2h_1+\gamma_1\kappa_1 )\int_{\mathbb{R}}\eta_1^2  \, dx \nonumber \\
&+ \varepsilon^2\kappa  \int_{\mathbb{R}}  \tilde{u} \eta \, dx  +\varepsilon^2(\kappa_1+\gamma_1h_1) \int_{\mathbb{R}}  \tilde{u}_1 \eta_{1} \, dx   \nonumber \\
&+\varepsilon^2 \delta^2\frac{1}{2}\int_{\mathbb{R}}\left(-\frac{h^2}{3\rho^2}(\rho h+3\rho_1 h_1)\tilde{u}_x^2 -\frac{h}{3\rho^2}(2\rho h^2 +6\rho_1h h_1+3\rho h_1)\tilde{u}_x \tilde{u}_{1,x}\right) \, dx \nonumber \\
&+\varepsilon^2 \delta^2\frac{1}{2}\int_{\mathbb{R}}\left(-\frac{1}{3\rho^2\rho_1}(\rho^2 h_1^3+\rho \rho_1h^3+3\rho \rho_1 h h_1^2 +3\rho_1h^2 h_1) \tilde{u}_{1,x}^2\right) \, dx . \nonumber \\
                   \label{H2Kdv}
\end{align}
Here $H^{(2)}$ is given in terms of $\tilde{u},  \eta, \tilde{u}_1,  \eta_1$ which are not canonical variables. The canonical are the variables
 
\begin{equation} \label{canvar} z=\xi+\frac{\Gamma}{2}\partial_x^{-1} \eta, \qquad z_1=\xi_1+\frac{\Gamma_1}{2}\partial_x^{-1} \eta_1. \end{equation} 
It is more convenient to work however with the variables which are $x$-derivatives of $z$ and $z_1$:
\begin{equation} \label{canvar2} p=\tilde{u}+\frac{\Gamma}{2} \eta, \qquad p_1=\tilde{u}_1+\frac{\Gamma_1}{2}\eta_1 .\end{equation}

In terms of these new variables, the Hamiltonian structure changes accordingly, \begin{equation} \label{PBKdV}
\frac{\mathrm{d}}{\mathrm{d}t}\begin{pmatrix} p\\ p_1\\ \eta\\ \eta_1 \end{pmatrix}=-\varepsilon^{-2}\partial_x \begin{pmatrix} 0 & 0 & 1 & 0\\0 & 0 & 0 & 1\\1 & 0 & 0 & 0\\0 & 1 & 0 & 0\end{pmatrix}\begin{pmatrix} \delta H/\delta p \\  \delta H/\delta p_1\\\delta H/\delta \eta \\ \delta H/\delta \eta_1 \end{pmatrix},
\end{equation}
In terms of the old variables  $\tilde{u},  \eta, \tilde{u}_1,  \eta_1$

\begin{equation} \label{PBKdV}
\frac{\mathrm{d}}{\mathrm{d}t}\begin{pmatrix} \tilde{u}\\ \tilde{u}_1\\ \eta\\ \eta_1 \end{pmatrix}=-\varepsilon^{-2}\partial_x\begin{pmatrix} -\Gamma & 0 & 1 & 0\\0 & -\Gamma_1 & 0 & 1\\1 & 0 & 0 & 0\\0 & 1 & 0 & 0\end{pmatrix}\begin{pmatrix} \delta H/\delta \tilde{u} \\  \delta H/\delta \tilde{u}_1\\\delta H/\delta \eta \\ \delta H/\delta \eta_1 \end{pmatrix},
\end{equation}

Since $\mathcal{O}(\varepsilon^2 \delta^2)=\varepsilon^3$ then already $H^{(2)}$ in the long wave approximation produces terms of order $\varepsilon^3$. For this reason we will not need terms of order  $\varepsilon^3 \delta^2 \sim\varepsilon^4$ from $H^{(3)}$. Therefore, the relevant terms  from $H^{(3)}$ are 

\begin{align}
H^{(3)}&=\varepsilon^3\frac{1}{2}\int_{\mathbb{R}}\left(\frac{1}{\rho}\eta\tilde{u}^2
+ \frac{2}{\rho}\eta \tilde{u}\tilde{u}_1 -\frac{\rho - \rho_1}{\rho \rho_1}\eta \tilde{u}_1^2 +\frac{1}{\rho_1}\eta_1 \tilde{u}_1^2 \right) \, dx \nonumber \\
&+\varepsilon^3\frac{1}{2}(\rho-\rho_1)\gamma^2\int_{\mathbb{R}}\frac{\eta^3}{3} \, dx    + \varepsilon^3\frac{1}{2}\rho_1\gamma_1^2\int_{\mathbb{R}}\frac{\eta_1^3}{3} \, dx      \nonumber \\
&+\varepsilon^3\frac{1}{2}\gamma\int_{\mathbb{R}} \tilde{u} \eta^2 \, dx    +\varepsilon^3\frac{1}{2}\gamma_1\int_{\mathbb{R}} \tilde{u}_1 \eta_1^2 \, dx  . \nonumber \\
             \label{H3Kdv}
\end{align}

The next assumption in our approximation is that the pair of canonical Hamiltonian variables $p_1$ and $\eta_1$ (i.e. with respect to the Hamiltonian structure in \eqref{PBKdV}) associated to the free surface are of smaller order, \begin{equation} \eta_1 \rightarrow \varepsilon \eta, \qquad p_1\rightarrow\varepsilon p_1.
\end{equation} 
Clearly $\tilde{u}_1\rightarrow\varepsilon\tilde{u}_1$. The approximate Hamiltonian of the system is with terms of orders  $\varepsilon^2$ and $\varepsilon^3$:

\begin{align}
H_a&=\varepsilon^2\frac{1}{2}\int_{\mathbb{R}}\left(\frac{h}{\rho}\tilde{u}^2
+ \varepsilon\frac{2h}{\rho}\tilde{u}\tilde{u}_1 + (\rho-\rho_1)(g+\gamma \kappa)\eta^2 \right) \, dx \nonumber \\
&+ \varepsilon^2\kappa  \int_{\mathbb{R}}  \tilde{u} \eta \, dx    -\varepsilon^2 \delta^2\frac{1}{2}\int_{\mathbb{R}}\frac{h^2}{3\rho^2}(\rho h+3\rho_1 h_1)\tilde{u}_x^2  \, dx \nonumber \\
&+\varepsilon^3\frac{1}{2}\int_{\mathbb{R}}\left(\frac{1}{\rho}\eta\tilde{u}^2
 \right) \, dx +\varepsilon^3\frac{1}{2}(\rho-\rho_1)\gamma^2\int_{\mathbb{R}}\frac{\eta^3}{3} \, dx   +\varepsilon^3\frac{1}{2}\gamma\int_{\mathbb{R}} \tilde{u} \eta^2 \, dx.    \nonumber \\
                   \label{HaKdv}
\end{align}

For the sake of simplification, let us introduce the following notations for some constants:
\begin{equation}\label{ak}
\begin{split}
a_1&=(\rho-\rho_1)(g+\gamma \kappa), \\
a_2&=\frac{h^2}{3\rho^2}(\rho h+3\rho_1 h_1) .
  \end{split}            
\end{equation}
The equations are

\begin{align}
\eta_t&=-\varepsilon^{-2}\left(\frac{\delta H_a}{\delta \tilde{u}}\right)_x, \nonumber \\
(\tilde{u}+\Gamma\eta)_t&=-\varepsilon^{-2}\left(\frac{\delta H_a}{\delta \eta}\right)_x ,\nonumber \\
\eta_{1,t}&=-\varepsilon^{-2}\left(\frac{\delta H_a}{\delta \tilde{u}_1}\right)_x, \nonumber \\
(\tilde{u}_1+\Gamma_1\eta_1)_t&=-\varepsilon^{-2}\left(\frac{\delta H_a}{\delta \eta_1}\right)_x=0. \nonumber \\                   \label{Haeq}
\end{align}

Due to the last equation we have
\begin{equation} 
\tilde{u}_1=-\Gamma_1\eta_1
\end{equation}

and for the other variables 
\begin{align} 
\eta_t&=-\left(\frac{h}{\rho}\tilde{u}+\kappa \eta +\varepsilon \frac{h}{\rho}\tilde{u}_1+
\delta^2a_2\tilde{u}_{xx}+\varepsilon\left(\frac{1}{\rho}\eta \tilde{u} +\frac{\gamma}{2}\eta^2\right)\right)_x, \label{eta-t} \\
(\tilde{u}+\Gamma\eta)_t&=-\left(a_1\eta+\kappa\tilde{u} \right)_x  -  \varepsilon \left( \frac{1}{2\rho}\tilde{u}^2+\frac{(\rho-\rho_1)\gamma^2}{2}\eta^2+\gamma \tilde{u} \eta \right)_x ,\label{u-t} \\
\eta_{1,t}&=-\varepsilon \frac{h}{\rho}\tilde{u}_x . \label{eta1}
\end{align}

The leading order linear equations for $\tilde{u}$ and $\eta$ are

\begin{equation}
\begin{split} 
\eta_t&=-\left(\frac{h}{\rho}\tilde{u}_x+\kappa \eta_x \right) , \\
(\tilde{u}+\Gamma\eta)_t&=-(a_1\eta_x+\kappa\tilde{u}_x) .
\end{split} \label{linear}
\end{equation}

The wavespeed $c$ of the solutions, proportional to $e^{ik(x-ct)}$ satisfies the quadratic equation

\begin{align} 
(c-\kappa)^2+\frac{h\Gamma}{\rho}(c-\kappa)+\frac{h}{\rho}(\kappa \Gamma - a_1)=0. \label{quad}
\end{align}

\noindent Introducing $c_0(k)=c(k)-\kappa $ and noting that $$ a_1-\kappa \Gamma=(\rho-\rho_1)(g-2\omega \kappa) \approx g(\rho-\rho_1)$$ we write the equation for $c_0(k)$ in the form

\begin{align} 
c_0^2+\frac{h\Gamma}{\rho}c_0-\frac{\rho-\rho_1}{\rho} g h =0. \label{quad}
\end{align}

\noindent The solution is

\begin{align} 
c_0=\frac{1}{2} \left(-\frac{h\Gamma}{\rho} \pm \sqrt{\left(\frac{h\Gamma}{\rho} \right)^2+4\frac{\rho-\rho_1}{\rho} g h} \right).
\end{align}

\noindent There are right ($c_0>0$) and left ($c_0<0$) running waves. We notice that in this approximation $c(k)$ is $k$-independent, i.e. constant. 

In the zero vorticity case 
\begin{align} 
c\to c'=\kappa \pm \sqrt{\frac{\rho-\rho_1}{\rho}gh} .
\end{align}

Let us introduce also $ c'_0=c'-\kappa. $  From \eqref{linear} it also follows that in the leading order $\tilde{u}=(\rho c_0 \eta)/h$, i.e.

\begin{align} \label{u-eta}
\tilde{u}=\frac{\rho c_0}{h}\eta + \varepsilon q +\mathcal{O} (\varepsilon^2)
\end{align}

\noindent for some yet unknown quantity $q$. Nevertheless from \eqref{eta1} and \eqref{u-eta} we determine

\begin{align} \label{u1-eta1}
\eta_1&=\varepsilon\frac{c_0}{c} \eta + \mathcal{O} (\varepsilon^2), \\
\tilde{u}_1& =-\Gamma_1 \eta_1=-\varepsilon\frac{\Gamma_1c_0}{c}\eta  + \mathcal{O} (\varepsilon^2).
\end{align}

Note that $\mathcal{O}(\tilde{u}_1)=\varepsilon$, thus $ \mathcal{O}(\varepsilon \tilde{u}_1)=\varepsilon ^2$ and such terms will be neglected in \eqref{eta-t} and \eqref{u-t} .  The most general form for $q$ is

\begin{align} \label{q}
\varepsilon q =  \varepsilon b_1 \eta^2 + \delta^2 b_2 \eta_{xx} + \mathcal{O} (\varepsilon^2)
\end{align}

\noindent for some constants $b_1, b_2.$ Now we are in a position to express everything in \eqref{eta-t} and \eqref{u-t} only via the variable $\eta$. This way we obtain two evolutionary equations for $\eta$ which should coincide up to $\mathcal{O}(\varepsilon)$. The equality of their coefficients allows us to find
\begin{align} \label{b-i}
b_1&=  \frac{\rho c_0(\gamma-\frac{c_0}{h})-2h\omega(\rho -\rho_1 )\gamma-2\Gamma c_0}{2h\left (2c_0+\frac{h\Gamma}{\rho}\right)},\\
b_2&=  \frac{ - a_2 \rho^2  c_0(\rho c_0+h\Gamma) }{h^2\left (2\rho c_0+h\Gamma\right)}.
\end{align}

\noindent $\eta$ satisfies the KdV equation \cite{KdV}, (see also \cite{ACbook,J02})

\begin{align} \label{KdVeq}
\eta_t+ c \eta _x + \delta^2 \left(\frac{h}{\rho}b_2+\frac{a_2 \rho c_0}{h} \right)\eta_{xxx} +\varepsilon \left(\frac{h}{\rho}b_1+\frac{ c_0}{h} +\frac{\gamma}{2}\right)2 \eta \eta_x=0  
\end{align}

\noindent which, in the limit when all vorticities are zero  becomes

\begin{align} \label{KdVlim}
\eta_t+ c' \eta _x + \delta^2 \frac{\rho c'_0 h}{6 \rho}(\rho h + 3 \rho_1 h_1) \eta_{xxx} +\varepsilon \frac{3 c'_0}{2h} \eta \eta_x=0.
\end{align}

\noindent From $\eta $ one can recover $\tilde{u}$ and as well as $\eta_1$  and $\tilde{u}_1$:

\begin{align} \label{u}
\tilde{u}&=\frac{\rho c_0}{h}\eta +  \varepsilon b_2 \eta^2 + \delta^2 b_3 \eta_{xx} + \mathcal{O} (\varepsilon^2), \\
\eta_1&=\varepsilon\frac{c_0}{c} \eta + \mathcal{O} (\varepsilon^2), \\
\tilde{u}_1& =-\varepsilon\frac{\Gamma_1c_0}{c}\eta + \mathcal{O} (\varepsilon^2).
\end{align}

\noindent We notice that in leading order $$\frac{\eta_1}{\eta} =\varepsilon \frac{c_0}{c}\sim \frac{c_0}{c_0+\kappa} $$ and therefore both positive and negative relative polarities for $\eta$ and $\eta_1$ are possible.
The KdV approximation for an internal wave coupled to a free surface for a different configuration of the currents is derived in \cite{CI}. 

The KdV equation represents a balance between a nonlinearity term $\varepsilon\eta \eta_x$, and dispersion term $\delta^2\eta_{xxx}.$ In the above considerations $\varepsilon$ and $\delta^2,$ are of the same order and as a result we can have the stable soliton solutions of the KdV equation. However, there are various geophysical scales and other situations are possible, including $\delta \sim \varepsilon^2.$ In such case $\delta^2 \sim \varepsilon^4 \ll \varepsilon$ and instead of the KdV equation the relevant model is the dispersionless Burgers equation ($\partial_{\tau}=\partial_t+c \partial_x$)

\begin{equation} \label{B} 
\eta_{\tau} +\varepsilon \left(\frac{h}{\rho}b_1+\frac{ c_0}{h} +\frac{\gamma}{2}\right)2 \eta \eta_x=0.  
\end{equation}

It is well known that the solutions of this equation always form a vertical slope and break. Such wave-breaking phenomenon is well known for internal waves in the ocean. This is a mechanism that causes mixing in the deep ocean, \cite{L}.

There are other integrable systems which provide an approximation of the equations in the Boussinesq regime, such as the Kaup-Boussinesq system
investigated firstly by D.J. Kaup \cite{K76}, see also \cite{IL12}. Two-component integrable systems, that can match the model equations up to order $\delta^2$, are the 2-component Camassa-Holm system and the Zakharov-Ito system \cite{CI08,I09,HI,FGL,EHKL,ELY}.

\section{Discussion and conclusions}

We consider a two-media system of liquids with different densities, free surface and a free internal surface separating the liquids. The bottom of the system is considered horizontal and flat. We studied the surface and internal waves driven by gravity and Coriolis forces and interacting with a current. The underlying current is in the form of a shear flow with a specific velocity profile. The current has constant vorticity at the horizontal strips where the surface and internal waves are localised. The model is aimed at geophysical applications, where a typical configuration is the one of a thin shallow layer of warm and less dense water over a much deeper layer of cold denser water. The governing equations are written in a canonical Hamiltonian form, which gives rise to a systematic approach for possible approximations. In particular, small amplitude and long-wave regimes are studied. There are various geophysical scales, allowing for smooth solitons at the KdV regime as well as breaking waves in the very large wavelength regime, when the equations are asymptotically equivalent to the dispersionless Burgers equation. In the case of a free surface, even in the case of very small amplitudes, the internal wave is usually coupled to the surface wave. This has an impact on the propagation speeds and is observed in other related models \cite{ConstIvMartin,CI}. Other asymptotic regimes, e.g. related to the Nonlinear Schr\"odinger equation \cite{TKM} remain to be studied.  The stability of the waves interacting with currents is another important issue that needs to be addressed in the future. Related recent works in this connection are \cite{CG,GH,DJH}.


\section*{Acknowledgements}
The author acknowledges Seed funding grant support from Dublin Institute of Technology for a project in association with the Environmental Sustainability and Health Institute,  Dublin.
The author is grateful to Prof. Adrian Constantin for many valuable discussions and to an anonymous referee for many important suggestions.





\end{document}